\documentclass[aps,prb,twocolumn,showpacs,showkeys,superscriptaddress]{revtex4-1}

\usepackage{graphicx}
\usepackage{dcolumn}
\usepackage{bm}
\usepackage{color}
\usepackage{hyperref}

\begin{document}

\preprint{APS/123-QED}

\title{Vacancy clusters at domain boundaries and band bending at the SrTiO$_3$(110) surface}

\author{Zhiming Wang}
\author{Xianfeng Hao}
\author{Stefan Gerhold}
\author{Michael Schmid}
\affiliation{Institute of Applied Physics, Vienna University of Technology, Vienna, Austria.}
\author{Cesare Franchini}
\affiliation{Faculty of Physics and Center for Computational Materials Science, University of Vienna, Vienna, Austria.}
\author{Ulrike Diebold}
\affiliation{Institute of Applied Physics, Vienna University of Technology, Vienna, Austria.}


\begin{abstract}
Antiphase domain boundaries (APDBs) in the ($n\,\times\,1$) ($n\,=\,4, 5$) reconstructions of the SrTiO$_3$(110) surface were studied with scanning tunneling microscopy (STM), x-ray photoemission spectroscopy (XPS) and density functional theory (DFT) calculations. Two types of APDBs form on each reconstruction; they consist of Ti$_x$O$_y$ vacancy clusters with a specific stoichiometry. The presence of these clusters is controlled by the oxygen pressure during annealing. The structural models of the vacancy clusters are resolved with DFT, which also shows that  their relative stability depends on the chemical potential of oxygen. The surface band bending can be tuned by controlling the vacancy clusters at the domain boundaries. 
\end{abstract}


\maketitle


\section{Introduction}
Strontium titanate (SrTiO$_3$), the archetypical perovskite oxide, has attracted intense attention for use in photocatalysis, microelectronics, and the emerging field of oxide electronics \cite{Mavroides:apl76,Wrighton:jacs76,McKee:sc01,Ohtomo:nat04,Mannhart:sc10}. These applications rely on a thorough understanding and control of the SrTiO$_3$ surface at the atomic scale. The possibility of nonstoichiometry, polarity, and mixed valences of metal cations make this difficult in such a multi-component system \cite{Noguera:rpp08,Marshall:prl11,Guisinger:acsnano09,Bonnell:rpp08}.  The SrTiO$_3$(001), for example, forms a variety of reconstructions depending on treatment parameters such as temperature, oxygen pressure, and Sr/Ti composition ratio near the surface \cite{Bonnell:rpp08,Shimizu:apl12,Gerhold:ss13}. In contrast, the SrTiO$_3$(110) surface has evolved as a unique system, where the surface structure is determined by the stoichiometry in the near-surface region and the transition between different surface reconstructions can be changed reversibly by depositing Sr or Ti, followed by annealing in O$_2$ \cite{Wang:prb11,Enterkin:natm10}. Here we show that, within particular surface phases, the  defect structure, and thus the surface potential, can be adjusted through the O$_2$ chemical potential.

Along the [110] direction a SrTiO$_3$ crystal is composed of alternately stacked (SrTiO)$^{4+}$ and (O$_2$)$^{4-}$ planes. As this results in a macroscopic dipole moment and infinite surface energy, the (110) surface is polar. This polarity is compensated by forming a homologous series of ($n$~$\times$~1) ($n$\,=\,3\,-\,6) reconstructions \cite{Enterkin:natm10}. These structures consist of corner-shared TiO$_4$ tetrahedra, forming networks of variable-membered rings. Specifically, six- and ten-membered rings have been found on the (4~$\times$~1) reconstruction \cite{Enterkin:natm10}, while the (5~$\times$~1) phase consists of eight- and ten-membered rings (see the structural model in Figure\,\ref{Model}, below) \cite{Enterkin:natm10,Li:prl11}. While SrTiO$_3$(110)  is symmetric along the [1$\overline 1$0] direction, these reconstructions are not.  As a consequence, antiphase domain boundaries (APDBs) form within the each one-dimensional (4~$\times$~1)/(5~$\times$~1) stripe. In previous theoretical and experimental studies, we have reported that single point defects, \textit{i.e.}, oxygen vacancies, do not form on the reconstructed surfaces \cite{Li:prl11,Wang:jpcc13,Wang:pnas14}.  However,  defect pairs of Ti$_2$O$_3$ vacancy clusters (here referred to as "type-I") and Sr adatoms decorate APDBs to relieve the surface stress and compensate the polarity, respectively \cite{Wang:prl13}.

In this paper, we report that a so-called type-II vacancy cluster, Ti$_4$O$_5$, can alternatively be formed at the (4~$\times$~1) APDBs. Correspondingly, two types of vacancy clusters, Ti$_3$O$_4$ (type-I) and Ti$_4$O$_5$ (type-II) are found at the APDBs of the (5~$\times$~1) surface. The type of APDB on each surface depends on the chemical potential of oxygen. In conjunction with density functional theory (DFT) calculations, experimental results demonstrate that control of the APDBs can be used to alter the surface band bending.

\section{Methods}
\subsection{Experiments}
The experiments were performed in a SPECS UHV system equipped with x-ray photoelectron spectroscopy (XPS) and STM with a base pressure of 1$\times$10$^{-10}$ mbar \cite{Wang:jpcc13}. Nb-doped (0.5 wt$\%$) SrTiO$_3$ single crystals (5\,mm~$\times$~5\,mm~$\times$~0.5\,mm) were purchased from MaTeck, Germany. A clean surface was prepared by cycles of Ar$^+$ sputtering (1\,keV, 5\,$\mu$A, 10\,minutes) followed by annealing in O$_2$ at pressures varying from 10$^{-7}$ to 10$^{-5}$ mbar at 1000 $^{\circ}$C for 1 h \cite{Wang:apl09}. In addition to STM, low-energy electron diffraction (LEED) was used to check for the uniformity of reconstructed places across the sample. The samples were heated by electron bombardment (13 mA, 900 V) at the back and the temperature was monitored with an infrared pyrometer. In XPS measurements, Mg $K\alpha$ radiation and a pass energy of 20 eV were used.

\subsection{Theory}
DFT calculations were carried out with the ``Vienna ab-initio simulation package'' (VASP) code \cite{vasp1,vasp2}. We have adopted the projector augmented-wave method  and the Perdew-Burke-Ernzerhof functional with a kinetic energy cutoff of 600 eV for plane waves and a single $k$ point ($\Gamma$ ) \cite{Blochl:prb94,pbe}. The surface structure was modeled with a supercell that was symmetric along the [110] direction, and consisted of a 9-layer slab separated by a vacuum layer of 12~\AA{}. The atoms in the central three layers were fixed, and the other atoms were allowed to relax until the force on each atom was less than 0.02 eV/\AA{}. Simulated STM images were generated with the Tersoff-Hamann approximation by integrating the local density of states from Fermi level (E$_F$) to 1.5 eV above the conduction band edge \cite{Tersoff:prl83}.

\begin {figure}[t]
 \includegraphics [width=3.4 in,clip] {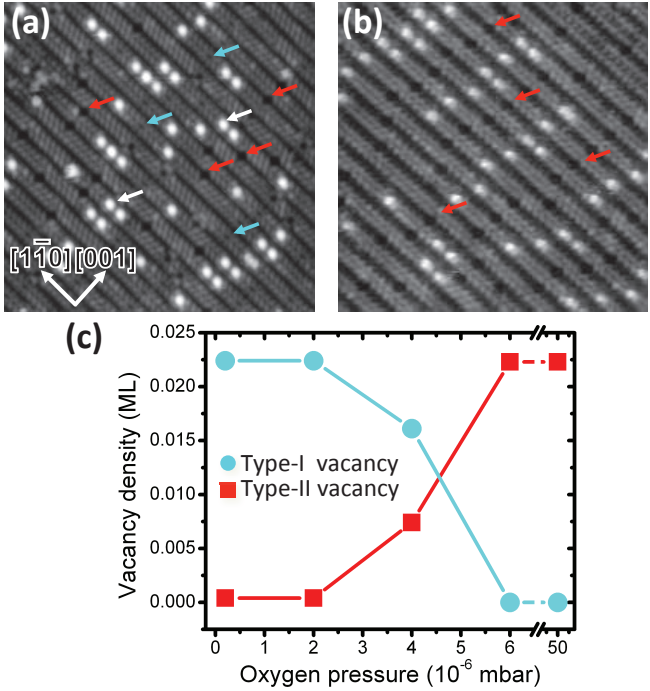}
 \caption{
 Control of the two types of domain boundaries at the SrTiO$_3$(110)-(4~$\times$~1) surface through annealing in O$_2$.   (a) STM image (20~$\times$~20 nm$^2$, V$_{sample}$ = +2.3 V, I$_{tunnel}$ = 0.3 nA). The type-I, type-II vacancy clusters and Sr adatoms are marked with cyan, red and white arrows, respectively. (b) STM image (20~$\times$~20 nm$^2$, V$_{sample}$ = +2.4 V, I$_{tunnel}$ = 0.3 nA) of a surface with only type-II vacancy clusters. (c)  Density of the vacancy clusters vs. O$_2$ pressure during annealing 1000\,$^{\circ}$C.
}
\label{APDB4x1}
\end{figure}

\section{Results} 

\subsection{SrTiO$_3$(110)-(4~$\times$~1) surface}

Figure\ \ref{APDB4x1}(a) shows an STM image of the SrTiO$_3$(110)-(4~$\times$~1) surface obtained by Ar$^+$ sputtering and annealing in 4~$\times$~10$^{-6}$ mbar O$_2$. The (4~$\times$~1) reconstructed surface shows quasi-one dimensional stripes along the [1$\overline{1}$0]  direction, corresponding to the Ti atoms in the six-membered rings \cite{Enterkin:natm10,Li:prl11,Wang:prl13}. Bright Sr adatoms (labeled by white arrows) and dark vacancy clusters (type-I, labeled by cyan arrows) are located at the APDBs \cite{Wang:prl13} at the middle of the stripes. In addition to these two features, larger vacancy clusters (type-II, labeled by red arrows) are observed at the trenches between two adjacent stripes. These larger vacancy clusters and Sr adatoms are quasi-ordered on the surface, with a similar spatial distribution as observed on the type-I surface \cite{Wang:prl13}. The relative ratio of these two types of vacancy clusters can be changed reproducibly by adjusting the oxygen pressure during annealing; type-I vacancy clusters can be obtained when annealing in ultra high vacuum (UHV) to low 10$^{-6}$ mbar oxygen pressure \cite{Wang:apl09,Wang:prb11,Wang:apl13,Wang:prl13,Wang:jpcc13}, while type-II vacancy clusters are found when annealing in a higher oxygen pressure [see Fig.\ \ref{APDB4x1}(c)]. This way, a mono-phased surface with type-II vacancy clusters is obtained [see Fig.\ \ref{APDB4x1}(b)]. The saturation density of both vacancy clusters is $\sim$~0.02 ML [here 1 ML~=~4.64~$\times$~10$^{14}$ atoms/cm$^2$ is defined as 1 atom per (1~$\times$~1) unit cell of bulk SrTiO$_3$(110) surface.] 

\subsection{SrTiO$_3$(110)-(5~$\times$~1) surface}

\begin {figure}[t]
 \includegraphics [width=3.4 in,clip] {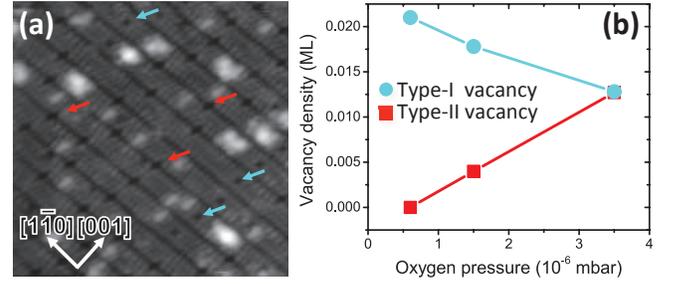}
 \caption{
Control of the two types of domain boundaries at the SrTiO$_3$(110)-(5~$\times$~1) surface through annealing in O$_2$.   (a) STM image (20~$\times$~20 nm$^2$, V$_{sample}$ = +2.2 V, I$_{tunnel}$ = 0.3 nA). The type-I and type-II vacancy clusters are marked with cyan and red arrows, respectively. (b) The density of the vacancy clusters depends on the O$_2$ pressure during annealing. 
}
\label{APDB5x1}
\end{figure}

\begin {figure}[b]
 \includegraphics [width=3.4 in,clip] {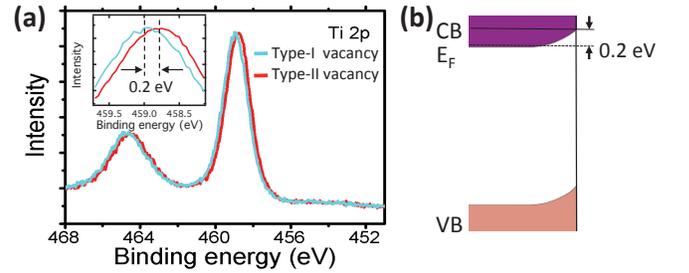}
 \caption{
(a) XPS core-level Ti \textit{2p} spectra of the surfaces with type-I (cyan curve) and type-II (red curve) vacancy clusters at domain boundaries. The inset shows that the spectra are shifted by about 0.2 eV. (b) Schematic diagram of the upward surface band bending. 
}
\label{BB}
\end{figure}

\begin {figure*}[t]
 \includegraphics [width=6.4 in,clip] {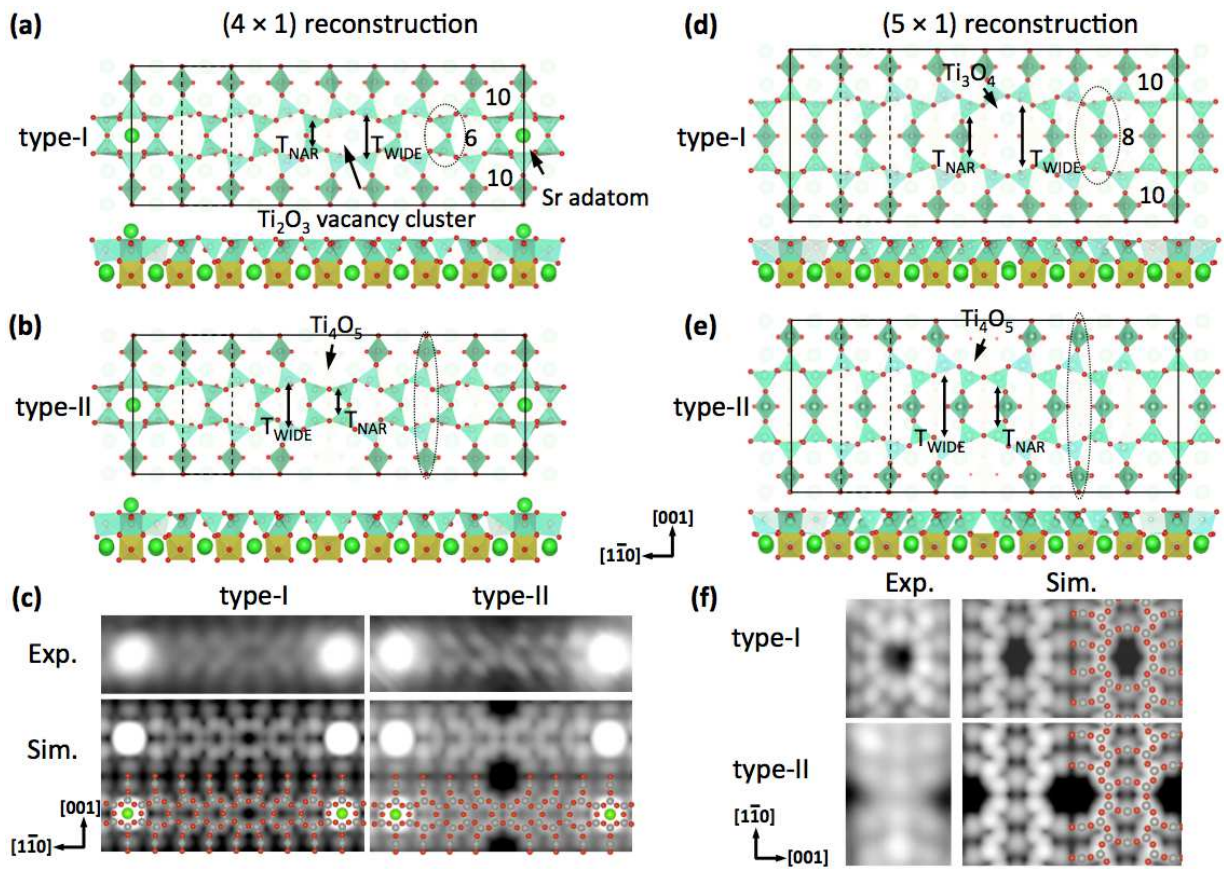}
 \caption{
Structural models for the reconstructed SrTiO$_3$(110) surface with antiphase domain boundaries (APDBs).  The left and right side shows APDBs on the (4~$\times$~1) and (5~$\times$~1) surfaces, respectively. The (4~$\times$~8)/(5~$\times$~8) and (4~$\times$~1)/(5~$\times$~1) cells are marked by full and dashed lines, respectively. Depending on whether the antiphase domains are connecting through wide (T$\rm{_{WIDE}}$) or narrow (T$\rm{_{NAR}}$) pairs of tetrahedra, type-I (a,c) and type-II (b, e) vacancy clusters are formed by removing the atoms enclosed in the dashed ellipsoids. In type-I domain boundaries, Sr$^{2+}$ adatoms compensate the charges introduced by the vacancy cluster.  (c, f) Corresponding simulated and experimental STM images. The structural models are superimposed on the simulated images \cite{Wang:prl13}.
}
\label{Model}
\end{figure*}

The (5~$\times$~1) reconstructed surface was obtained by depositing $\sim$\,0.15 ML [1\,ML\,=\,4.64\,$\times$\,10$^{14}$\,atoms/cm$^2$ relative to SrTiO$_3$(110) surface] Sr metal on the (4~$\times$~1) surface followed by annealing  \cite{Wang:prb11}. The (5~$\times$~1) surface exhibits stripes consisting of three bright rows along the [1$\overline{1}$0]  direction. Figure\,\ref{APDB5x1}(a) shows an STM image of the SrTiO$_3$-(5~$\times$~1) surface obtained after annealing in 4~$\times$~10$^{-6}$ mbar O$_2$. Again two types of vacancy clusters coexist on the surface. The type-I vacancy clusters appear at the center of the stripes [labeled by cyan arrows in Fig.\,\ref{APDB5x1}(a)], whereas the type-II vacancy clusters appear at the trenches between the (5~$\times$~1) stripes [labeled by red arrows in Fig.\,\ref{APDB5x1}(a)]. On the (5~$\times$~1) surface the relative ratio of these two types of vacancy clusters is also regulated by the oxygen pressure during annealing [see Fig.\,\ref{APDB5x1}(b)]. 

\subsection{X-ray Photoelectron Spectroscopy}

The vacancy clusters on the APDBs affect the surface electronic properties. Figure\ \ref{BB}(a) shows Ti \emph{2p} core-level spectra of the surfaces with type-I and type-II vacancy clusters on the (4~$\times$~1) surface. No Ti$^{3+}$ states are detected on either surface. The spectra of the surface with type-II vacancy clusters shifts about 0.2 eV to lower binding energy as compared to the type-I vacancy surface [see inset in Fig.\ \ref{BB}(a)]. The same binding energy shift is also observed for Sr \emph{3d} and O \emph{1s} core-level spectra (not shown). We conclude that surface band bending occurs when the vacancy cluster type changes. Previous valence-band spectra showed that the SrTiO$_3$(110)-(4~$\times$~1) surface with type-I vacancy cluster has a negligible band bending \cite{Wang:prb11,Wang:jpcc13}. Therefore, upon formation of type-II vacancy clusters the bands bend upward relative to the Fermi level  as shown in the schematic diagram in Fig.\ \ref{BB}(b). A similar surface band bending is also observed on the (5~$\times$~1) surface upon formation of type-II vacancy clusters (not shown).

\section{Discussion}

\begin {figure}[t]
 \includegraphics [width=3.2 in,clip] {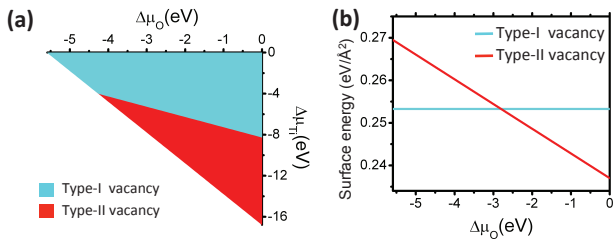}
 \caption{
(a) Thermodynamic stability of type-I and type-II vacancy clusters on the SrTiO$_3$(110)-(4~$\times$~1) surface in dependence on the relative Ti and O chemical potential. The dark, red area corresponds to an environment in where type-II vacancy cluster is stable while the light, cyan area corresponds to that in which type-I vacancy cluster is stable.Ê (b) The surface grand potential of type-I and type-II vacancy cluster as function of $\Delta\mu_\mathrm{O}$. For details see the Supplement.
}
\label{Stablility}
\end{figure}

\subsection{Structural model}

In order to understand the formation of the two types of vacancy clusters, the surface structure is analyzed. On the SrTiO$_3$(110)-(4~$\times$~1) surface the six-membered ring can be viewed as pairs of narrow (T$\rm{_{NAR}}$) and wide (T$\rm{_{WIDE}}$) tetrahedra [see Fig.\ \ref{Model}(a)]. Depending on the local atomic arrangement at the APDBs two types of vacancy clusters are formed. When two antiphase domains are connected through two T$\rm{_{WIDE}}$, the type-I vacancy cluster results [see upper part of Fig.\ \ref{Model}(a)]. This requires removing a pair of T$\rm{_{NAR}}$ with a Ti$_2$O$_3$ stoichiometry, and two nominal positive charges \cite{Wang:prl13}. A type-II vacancy cluster is formed when two T$\rm{_{NAR}}$ are linked together. Different structural models of the type-II APDB have been compared with DFT calculations. The optimized model is shown in Fig.\ \ref{Model}(b). In this case, in addition to a pair of T$\rm{_{WIDE}}$, two TiO$_4$ tetrahedra in the middle of the ten-membered ring are removed with an overall stoichiometry and nominal charge of (Ti$_4$O$_5$)$^{6+}$ [see Fig.\ \ref{Model}(b)]. Simulated STM images based on relaxed structures of all the vacancy clusters agree well with the experimental ones [see Fig.\ \ref{Model}(c)].

We can readily extend the structural models of APDBs to the (5~$\times$~1) surface. In correspondence with the (4~$\times$~1), the (5~$\times$~1) surface consists of eight- and ten-membered rings \cite{Li:prl11}, and similar narrow and wide tetrahedra pairs are present in the eight-membered rings (labeled as T$\rm{_{NAR}}$ and T$\rm{_{WIDE}}$ in Fig.\ \ref{Model}(d)). Forming the type-I vacancy cluster requires the removal of a pair of T$\rm{_{NAR}}$ and the middle TiO$_4$ in the eight-membered ring [labeled in Fig.\,\ref{Model}(d)] with an overall stoichiometry and nominal charge of (Ti$_3$O$_4$)$^{4+}$. The type-II vacancy cluster on the (5~$\times$~1) surface is the same as the one on the (4~$\times$~1) surface [see Fig.\ \ref{Model}(e)].

\subsection{Thermodynamic stability}

To further clarify the structural models and study their relative thermodynamic stability, we have calculated the surface grand potential of both types of vacancy clusters (see the Supplementary Information). Figure\ \ref{Stablility}(a) shows the relative thermodynamic stability of the surface with type-I and type-II vacancy clusters as function of $\Delta\mu_\mathrm{O}$ and $\Delta\mu_\mathrm{Ti}$. In an oxygen-poor environment ($\Delta\mu_\mathrm{O}<$~-4.2~eV) the type-I vacancy cluster is more stable than the type-II vacancy cluster. In an oxygen-rich and Ti-poor environment ($\Delta\mu_\mathrm{O}=$~0~eV, $\Delta\mu_\mathrm{Ti}<$~-8.3~eV) the type-II vacancy cluster becomes more stable. Thus, a phase transition can be induced by changing the oxygen and Ti chemical potentials.  

In order to evaluate the stability of the two vacancy types as a function of $\Delta\mu_{O}$ only, we assume the surface is also in equilibrium with TiO$_2$ \cite{Kienzle:prl11}, since the topmost reconstruction layer has a Ti valence of 4+ [see Fig.\ \ref{BB}(a)] \cite{Russell:2008prb,Wang:prb11,Wang:jpcc13}. Since the type-I surface can be expressed as stoichiometric SrTiO$_3$ and TiO$_2$, its surface energy will not depend on the oxygen chemical potential. The surface energy of type-II vacancy clusters, however, changes as a function of $\Delta\mu_{O}$ as shown in Fig.\ \ref{Stablility}(b). In an oxygen-poor ($\Delta\mu_{O}<-$2.8 eV) environment, the type-I vacancy cluster is more stable than the type-II vacancy cluster. The type-II vacancy cluster becomes more stable when $\Delta\mu_{O}$ increases, consistent with the experimental result shown in Fig.\ \ref{APDB4x1}(c).

\begin {figure}[t]
 \includegraphics [width=2.6 in,clip] {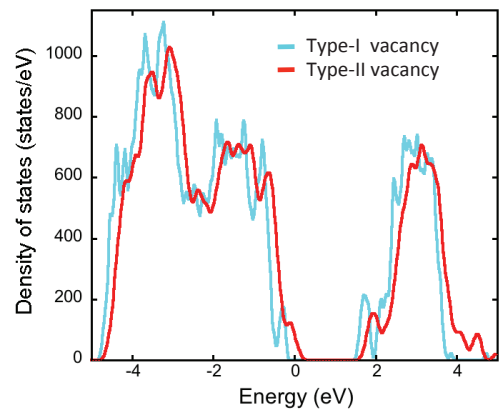}
 \caption{
DFT-calculated density of states of surface with type-I (cyan curve) and type-II (red curve) vacancy cluster on the domain boundaries of the SrTiO$_3$(110)-(4~$\times$~1) surface, respectively. 
}
\label{DOS4x1}
\end{figure}

\subsection{Electronic properties}

\begin {figure}[t]
 \includegraphics [width=2.6 in,clip] {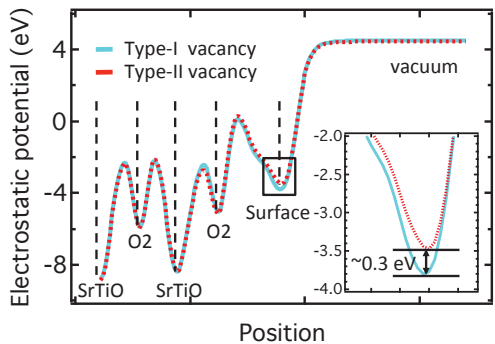}
 \caption{
DFT-calculated electrostatic potential of models with type-I and type-II vacancy clusters. The potential curves are a planar average of the electrostatic potential. The inset shows that a 0.3 eV potential difference appears at the topmost layer between different type vacancy cluster surface.
}
\label{EP}
\end{figure}

Several mechanisms can induce an upward surface band bending, such as healing of oxygen vacancies, Fermi level pinning, and adsorbing charged species. Sample self-reduction could occur during annealing \cite{Herranz:prl07}, but we exclude that the upward band bending originates from oxygen vacancies from the absence of Ti$^{3+}$ on both surfaces in the XPS spectra [see Fig.\ \ref{BB}(a)]. This is consistent with previous theoretical and experimental results that showed that the oxygen vacancy density is extremely low on this surface \cite{Li:prl11,Wang:jpcc13}. In addition, previous studies show that no in-gap states emerge from the formation of type-I vacancy clusters \cite{Wang:jpcc13,Wang:pnas14}. This is confirmed by DFT calculations that show both types vacancy cluster surfaces do not have in-gap states (see Fig.\ \ref{DOS4x1}). Thus, we can exclude that the upward band bending observed here is due to Fermi level pinning \cite{Zhang:cr12}. The surface band bending is due to the differently charged state of the vacancy clusters. 

The different types of vacancies explain the observed band bending. Here we focus on the (4~$\times$~1) surface. The (4~$\times$~1) reconstruction compensates the surface polarity of SrTiO$_3$(110) \cite{Enterkin:natm10}. Vacancy clusters alter the surface stoichiometry, thus  they also disturb (or disrupt/affect) the charge (and surface polarity) compensation of the  (4~$\times$~1) surface. As discussed above, each type-I vacancy cluster introduces two negative charges into the structure. It turns out that these are largely compensated by Sr adatoms, each with two positive charges \cite{Wang:prl13}. Defect pairs with a  type-I vacancy cluster and a Sr adatom [see Fig.\ \ref{Model}] have been found to be thermodynamically stable in both experiments and DFT calculations \cite{Wang:prl13}; thus no surface band bending occurs on the type-I surface \cite{Wang:prb11,Wang:jpcc13,Wang:pnas14}.

The type-II Ti$_4$O$_5$ vacancy clusters that form under O-rich conditions will (formally) introduce six negative charges. These can no longer be compensated by Sr adatoms - the required high Sr density is not thermodynamically stable, and a phase transition is induced upon annealing \cite{Li:prl11,Wang:prb11}.  Thus, upward band bending occurs when the negatively charged type-II vacancy clusters are formed \cite{Zhang:cr12}. The band bending is estimated by comparing the electrostatic potential of both models in DFT calculations (see Fig.\ \ref{EP}). In the central SrTiO$_3$ layers the potential is the same but a $\sim$0.3 eV difference appears at the topmost layer. This is in good agreement with the experimental observations. A similar band bending is observed on the (5~$\times$~1) surface and confirmed by theoretical calculations (not shown). Therefore, we conclude that the surface band bending can be induced by formation of  the Ti$_4$O$_5$ vacancy clusters at APDBs. 

Tuning the band bending on oxide surfaces by structural modification and/or external electric field is of importance in various applications. For example, band bending influences  photocatalytic processes \cite{Zhang:cr12}, as it affects the adsorption or desorption of reactants on the surface, as well as charge carrier recombination and transfer between the oxide and an adsorbate. Band bending has been used to tune the electronic properties of oxide surfaces and even interfaces by external electric fields and surface adsorbates \cite{Ueno:natm08,Xie:am13}. Engineering the band bending on the SrTiO$_3$(110) surface can provide a flexible approach to optimize the performance in photocatalysis and tune the properties of two-dimensional electron gases on the SrTiO$_3$(110) surface and interface \cite{Wang:pnas14,Annadi:natc13}.

\section{Conclusion}

In summary, we have demonstrated subtle control of APDBs in the ($n$~$\times$~1) ($n$ = 4,5) reconstructions of the SrTiO$_3$(110)  surface. Depending on the atomic configurations at APDBs,  two types of vacancy clusters are formed at each reconstruction. Their relative stabilities can be tuned by changing the oxygen chemical potentials; this, in turn, allows tuning the surface band bending through reversible and controllable switching of vacancy cluster types.\\

This work is supported by the Austrian Science Fund (FWF) under Project No. F45 and the ERC Advanced Research Grant ``OxideSurfaces''.

%
%



\providecommand{\WileyBibTextsc}{}
\let\textsc\WileyBibTextsc
\providecommand{\othercit}{}
\providecommand{\jr}[1]{#1}
\providecommand{\etal}{~et~al.}

\end{document}